\documentclass[prd,twocolumn]{revtex4}
\usepackage{graphicx}
\usepackage{amstext,amsmath,amssymb,amsfonts,bbm}
\usepackage{amsthm}
\usepackage{euscript}

\def\be{\begin{equation}}
\def\ee{\end{equation}}
\def\bes{\begin{eqnarray}}
\def\ees{\end{eqnarray}}

\newcommand{\C}{\mathbb{C}}

\def\eps{\epsilon}

\renewcommand{\S}{\sigma}

\def\arr{\rightarrow}

\def\pp{\partial}

\newcommand{\SU}{\mathrm{SU}}

\def\aa{{\cal A}}
\def\hh{{\cal H}}
\def\ii{{\cal I}}
\def\mm{{\cal M}}

\def\Int{{\rm Int}}

\def\6{\langle}
\def\9{\rangle}

\def\tr{{\rm tr}\,}

\def\inte{{\rm int}}

\def\half{\mbox{$1\over2$}}
\def\ha3{\mbox{$3\over2$}}
\def\Id{{\mathbbm 1}}

\newcommand{\ket}[1]{|#1\rangle}

\def\vio{{\vec{\ii}}}
\def\jj{{\cal J}}
\def\vjo{{\vec{\jj}}}

\begin{document}
\title{ Quantum information in loop quantum gravity}
\author{Daniel R. Terno}\email{dterno@perimeterinstitute.ca}
\affiliation{Perimeter Institute for Theoretical Physics,
31 Caroline St, Waterloo, Ontario, Canada N2L 2Y5}

\begin{abstract} A coarse-graining of spin networks is expressed
in terms of partial tracing, thus allowing to use tools of quantum
information theory. This is illustrated by the analysis of a
simple black hole model, where the logarithmic correction of the
Bekenstein-Hawking entropy is shown to be equal to the total
amount of correlations on the horizon. Finally other applications
of entanglement to quantum gravity are briefly discussed.
\end{abstract}
\maketitle
\section{Introduction}
 Loop Quantum Gravity (LQG) is a canonical quantization of
General Relativity, which relies on a 3+1 decomposition of
space-time
\cite{lqg2,lqg3}. It describes the states of 3d geometry
and their evolution in time (through the implementation of a
Hamiltonian constraint). The states of the canonical hypersurface
are the  {\it spin networks}, which represent polymeric
excitations of the gravitational field. Spin networks are also
used to describe the states in the path integral approach to the
quantization of gravity --- spin foams
\cite{spin,oriti}.

The main prediction in the LQG framework is a discrete spectrum of
geometric operators. In particular, a surface can be regarded as
made of elementary patches of a finite quantized area. Based on
this
 structure  one can not only study  the
entropy associated to a surface
\cite{lqg2,lqg3, bhlqg}, but also analyze the
information-theoretical aspects of the corresponding states
\cite{etflo,etd1,etd2}.

 In this work I describe the precise relationship between the coarse-graining of a spin network
 and partial tracing. This allows  to discuss entanglement in spin networks and
 to relate it to the black hole entropy.

\section{Spin networks, coarse graining and partial tracing}

A spin network is a graph $\Gamma$ with vertices $v$ and oriented
edges $e$. The spin network state is  the assignment of a $\SU(2)$
representation $V^{j_e}$ to each edge $e$ and a $\SU(2)$-invariant
linear map (an intertwiner) $
\ii_v:\,\bigotimes_{e\,{\rm ingoing}} V_{j_e}\arr \bigotimes_{e \,{\rm outgoing}} V_{j_e}
$
to each vertex $v$. Denote the Hilbert space of intertwiners at
the vertex $v$ as $\hh_v\equiv\Int(\bigotimes_{e\,{\rm ingoing}}
V_{j_e}\arr\bigotimes_{e\,{\rm outgoing}} V_{j_e})$.

A spin network state $|\Gamma,\vec{\jmath},\vio\9$  defines a
function of the holonomies along the graph edges
$T_{(\Gamma,\vec{\jmath},\vio)}[g]$. 
 For a fixed graph and
a fixed assignment of the representations we omit their labels and
denote a basis state in $\hh^0\equiv\otimes_v\hh^0_v$ as
$\ket{\ii_1\ldots
\ii_V}\equiv\ket{\vio\, }$, where each $\ii_v$ enumerates the
intertwiners at the vertex $v$. The corresponding function
$T_{\vio}[g]$  equals to the tensor contraction of the matrix
representations of the group elements $\bigotimes_eD^{j_e}(g_e)$
with the intertwiner $\bigotimes_v
\ii^{\iota_v}$,
\be
T_{\vio\ }[g]\equiv\6g|\ii_1\ldots \ii_V\9
\equiv\tr\bigotimes_{v=1}^V\bigotimes_{e=1}^{E}D^{j_{e_v}}(g_{e_v})\cdot
\ii_v,
\ee
where $E$ is a number of edges and $V$ a number of vertices of the
graph.
 It is a gauge invariant function, its value is preserved under the (residual) action
 of the SU(2) gauge group at the graph's vertices.
Such gauge invariant functions, called gauge invariant cylindrical
functions, are the wave functions of quantum geometry
\cite{lqg2,lqg3,spin}.

 A note on normalization:  the intertwiners are normalized as $\|\ii_v|=1$,
 and the spin network states are normalized to one $\vjo$,
$\6\vio~|\vjo\9=\delta_{\vio\, \vjo}$ by absorbing the factors
$\prod_e 1/\sqrt{d_{j_e}}$.
 An arbitrary pure state is given by $
|\Psi\9=\sum_{\vio }c_\vio\ket{\vio }$, with
$\sum_{\vio}|c_\vio|^2=1$.

Consider a closed connected spin network based on the oriented
graph $\Gamma$ and a bounded connected region $B$ of this spin
network. The interior ${\rm int}(B)$ of $B$ consists of the set of
vertices $v\in B$ and the edges between them. The exterior ${\rm
ext}(B)$ of $B$ consists of all other vertices. Its boundary $\pp
B$ consists of the set of edges $e$ such that one of its end
vertices is inside $B$ and the other is outside. The state of $B$
is the tensor product of all the intertwiners attached to the
vertices $v\in {\rm int}(B)$: $\hh_B=\hh_{{\rm int}(B)}\equiv
\bigotimes_{v\in B}\hh_v$. The state of ${\rm ext}(B)$ is the tensor product of all
the intertwiners attached to the vertices $v\in {\rm ext}(B)$:
$\hh_{{\rm ext}(B)}\equiv
\bigotimes_{v\notin B} \hh_v$. The Hilbert space of boundary
states $\hh_{\pp B}$ is the space of intertwiners between the
representations $j_e$ attached to the edges crossing the boundary
$\pp B$.

 In the simplest coarse-graining  procedure one contracts
the intertwiners attached to each internal vertex and thus obtains
an intertwiner
between the edges crossing the boundary $\pp B$. 
 One can  glue these same intertwiners
using a non-trivial parallel transport between the internal
vertices, i.e., using non-trivial group elements $g_e$ on each
internal edge $e\in {\rm int}(B)$. For an arbitrary set of group
elements $\{g_e, e\in E_B\}$,  the parallel-transport dependent
boundary state in $\hh_{\pp B}$ is
\cite{etd3}.\pagebreak
\begin{widetext}
\be
\ii_B[g^{(0)}_{e\in B}]\,\equiv\,
\int_{\SU(2)}dg\, \tr_{e\in B}\,\left[
\bigotimes_{e\in \pp B} D^{j_e}(g)^{\eps_e}
\bigotimes_{e\in B} D^{j_e}(g^{(0)}_{e})
\cdot\bigotimes_{v\in B}\ii_v.
\right],
\ee
\end{widetext}
 where $\eps_e, e\in E_B$ is a sign $\pm$
depending on whether the edge $e$ is ingoing  ($s(e)\notin B$) or
outgoing ($s(e)\in B$). The trace is taken over all the $\SU(2)$
representations $V_{j_e}$ labelling the edges that are linked to
the vertices of $B$. The integration over $\SU(2)$ ensures that
the resulting tensor is $\SU(2)$ invariant, thus an intertwiner
\cite{etd3}.

Due to the global $\SU(2)$ invariance and to the $\SU(2)$
invariance of the intertwiners $\ii_v$, not every distinct set of
group elements $\{g_e, e\in E_B\}\in \SU(2)^{E_B}$ leads
 to a distinct boundary state. To get distinct states one has to
quotient by the gauge invariance. The simplest orbit is the
one-point orbit defined by $g_e=\Id,\, \forall e\in E_B,$ which
corresponds to the contraction of the internal intertwiners
\cite{etd3}.

Recall now the standard definition of a reduced density operator
\cite{per,ll}. Consider two
subsystems $A$ and $B$,   $\hh=\hh_A\otimes\hh_B$, with the direct
product basis $\ket{mn}\equiv\ket{m}_A\otimes\ket{n}_B$. The
corresponding wave functions are $\6x|m\9=\psi_m(x)$ for $A$ and
$\6y|n\9=\phi_n(y)$ for $B$. A generic state $\ket{\Psi}\in\hh$ is
given by a linear combination $\ket{\Psi}=c_{mn}\ket{m}\ket{n}$.
If the matrix elements of the operator $O$ are
$\6mn|O|m'n'\9=o_{mm'}\delta_{nn'}$,
 then
\be
\6\Psi|O|\Psi\9=\sum_{m,m',n}o_{mm'}c_{mn}\bar{c}_{m'n}=\tr(o\rho_\Psi^A),\label{expred}
\ee
where the reduced density operator
$\rho_\Psi^A\equiv\tr_B\rho_\Psi$ is obtained by tracing out the
subsystem $B$, $\rho^A_{mm'}=\rho_{mn,m'n}$. In the coordinate
basis the operator $O$ is given by
\be
O(x,y;x',y')=o(x,x')\delta(y-y'), \label{matelc}
\ee
 so thanks to the orthonormality of the functions $\phi_n$ the reduced density
 operator is
\begin{align}
 \rho^A_\Psi(x,x')=\int dy
 dy'\Psi(x,y)\overline{\Psi}(x',y')\delta(y-y')\nonumber\\=
 \sum_{m,m',n}c_{mn}c_{m'n}\psi_m(x)\bar{\psi}_{m'}(x'),\label{rhoc}
\end{align}
with $ \6\Psi|O|\Psi\9=\int dx dx'o(x,x') \rho^A_\Psi(x,x')$.

 In
the language of the intertwiners, a partial tracing does not
present any new features. Taking an intertwiner
$|\vio\9=\otimes_{v\in\Gamma}|\ii_v\9$ and tracing out $B$
produces a basis state of $A$, $|\vio_A\9=\otimes_{v\in
A}|\ii_v\9\in\hh^0_A$, with a corresponding expressions for the
reduced density matrices of general states.

 Expectation values of the operators that depend
only on the region $A$ are calculated according to
Eq.~(\ref{expred}). For example, the volume operator \cite{lqg2,
lqg3} is a sum of the vertex operators. The isomorphism between
the spin network states and zero-angular momentum states of the
corresponding abstract spin system leads to  an expectation value
of the form of Eq.~(\ref{expred}). In particular, if $A$ contains
a single vertex $v=1$ and the operator's  matrix elements are
$(O_1)_{\vio\vjo}=o(\ii_1,\jj_1)\prod_{v=2}^V\delta_{\ii_i\jj_i}$,
then for a generic pure state
$|\Psi\9=\sum_{\vio}s_\vio|\vio\9\in\hh^0_\Gamma$ the expectation
is
\be
\6\Psi|O_1|\Psi\9=\sum_{\vio,\jj_1}o(\ii_1,\jj_1)s_{\ii_1\ii_2\ldots\ii_V}\bar{s}_{\jj_1\ii_2\ldots\ii_V}.\label{correct}
\ee

On the other hand, the total coarse-graining of $B$ results in
\be
\ii_B\equiv\tr_{v\in B}\!\!\bigotimes_{e\in\inte(B)}\!
D^{j_e}(\Id)\cdot\bigotimes_{v\in B}\ii_v\in\hh_{\pp B},
\ee
where $\tr_{v\in B}$ denotes the summation over pairs of indices
pertaining to the vertices in the region $B$. It turns a
normalized basis state
$|\vio_A\9\otimes|\vio_B\9\in\hh^0_\Gamma=\otimes_{v\in\Gamma}\Int_v$
into a non-normalized state (see \cite{etd3})
$|\ii_A\9\otimes|\vio_B\9\in\hh_{\pp A}\otimes\hh^0_B$. It can be
decomposed as $ |\ii_B\9=\sum_\alpha
c^{\vio_B}_\alpha|\ii_B^\alpha\9, $ where the intertwiners
$\ii_B^\alpha$, $\alpha=1,\ldots,\dim\hh_{\pp B}$, form the
orthonormal basis of $\hh_{\pp B}$. A general coarse-graining
procedure  leads to the same equation, but with the coefficients
depending on the holonomies $\{g_e\}_{e\in\inte(B)}$. A pure state
$|\Psi\9=\sum_{\vio}s_\vio|\vio\9\in\hh^0_\Gamma$ becomes a
 pure state
\be
|\Psi_{[B]}\9=\sum_{\vio_B,\vio_A,\alpha}s_{\vio_B\vio_A}c^{\vio_B}_\alpha|\ii_B^\alpha\9|\vio_A\9/\sqrt{\|\ii_B\|},
\ee
in $\hh_{\pp B}\otimes\hh_A^0$.

Cylindrical functions do not allow a natural separation of
variables in $O_1(g,g')$ that is analogous to Eq.~(\ref{matelc}),
since there is no relation between the number of the intertwiners
$\ii_v$, $v=1,\ldots,|V|$, that define the tensor product
structure of $\hh^0_\Gamma$, and the number of edges, that define
the structure of $H_\Gamma=L^2(\SU(2)^{E}/\SU(2)^{V})$. To obtain
a wave functional that corresponds to $|\vio_{A}\9$ one needs to
contract with the representations $D^{j_e}$ that corresponds to
all the edges of the coarse-grained graph $\Gamma[B]$. Hence the
coarse-graining procedures that were described above allow to
introduce the reduced subsystems in the language of cylindrical
functions. From the definition of a total coarse-graining of a
spin network state $\phi_\vio(g)=\6g|\Gamma,\vec{\jmath},\vio\9$
over a region $B$ results in $\phi_{\vio [B]}(g)\equiv\6
g|\vio_A\tilde{\ii}_B\9$.
 It is
given explicitly by
\be
\phi_{\vio [B]}(g)=\tr\!\!\! \bigotimes_{e\notin \inte(B)}\!\!D^{j_e}(g_e)\cdot
\mm,
\ee
with
\be
 \mm=\tr_{v\in B}\!\!\! \bigotimes_{e\in\inte(B)}\!\!\!
D^{j_e}(\Id)\cdot\bigotimes_{v\in\Gamma}\ii_v/\sqrt{\|\ii_B\|}.
\ee

Consider for simplicity a pure state $|\Psi\9\in\hh^0_\Gamma$.
Mixed states are treated analogously. From the above facts it
follows \cite{etd3} that  the coarse-grained states $\phi_{\vio
[B]}(g)$ play a role of the basis wave functions $\varphi_m(g)$ in
the standard partial tracing. In particular, a ``coordinate"
expression of the  reduced density matrix
$(\rho_\Psi^A)_{\vio_A\vjo_A}=\sum_{\vio_B}s_{\vio_A\vio_B}\bar{s}_{\vjo_A\vio_B}$
is given by
\begin{widetext}
\be
\rho_\Psi^A(g,g')=
\sum_{\vio,\
\vjo}s_{\vio}\overline{s}_{\vjo}\!\int
(dg)^{E_A}(dg')^{E_A}\!\prod_{e\in
\inte(B)}\!\delta(g_e,g'_e)\phi_{\vio
}(g)\overline{\phi}_{\vjo }(g')/\|\ii_B\|, \label{defred}
\ee
\end{widetext}
which is equivalent to
\be
\rho_\Psi^A(g,g')=\sum_{\vio,\
\vjo}s_{\vio_A\vio_B}\overline{s}_{\vjo_A\vio_B}\phi_{\vio[B]
}(g)\overline{\phi}_{\vjo[B] }(g'),\label{goal}
\ee
while the  reduced functional matrix elements of $(O_1)$ are
\be O_1^A(g,g')=\sum_{\vio, \jj_1}\phi_{\vio[B] }(g)
o(\ii_1,\jj_1)\overline{\phi}_{\vjo[B] }(g'),
\ee
where $|\vjo\9\equiv|\jj_1\ii_2\ldots\ii_V\9$.

\section{Black hole entropy}

A generic surface on a spin network background is thus described
as a set of patches, each punctured by a unique link of the spin
network.   
The spin network defines how the patches, and therefore the whole
surface, is embedded in the surrounding 3d space and describes how
the surface folds. For a closed surface, the region of the spin
network which is inside the surface defines an intertwiner between
the patches of the surface.

An important remark is that any spin $j$ representation $V^j$ can
be decomposed as a  symmetrized tensor product of $2j$
spin-\half{} representations $V^{1/2}$. Therefore, one can
interpret that a fundamental patch or elementary surface is a
spin-\half{} representation. All higher spin patches can be
constructed from such elementary patches. For example, considering
two spin-\half{} patches, they can form a spin 0 representation or
a spin 1 representation: in one case, the two patches are folded
 one on another and cancel each other, while in the later case they
add coherently to form a bigger patch of spin 1. Considering an
arbitrary surface, one can then look at it at the fundamental
level, decomposing it into spin-\half{} patches, or one can look
at it at a coarse-grained level decomposing the same surface into
bigger patches of spin $s> \half$. From this point of view, the
size of the patches used to study a surface is like the choice of
a ruler of fixed size used by the observer to analyze the
properties of the object. Thus  one can study the coarse-graining
or renormalisation of these quantities when one observes the
surface at a bigger scale, using bigger patches to characterize
the surface \cite{etd2}.

Considering the horizon as a closed surface the interior of the
black hole is described by  (a superposition of) spin networks
with their edges puncturing the horizon and defining the patches
of the horizon surface. For an external observer only the horizon
information is relevant.  For him  the bulk spin network is fully
coarse-grained and  the state of the $n$ patches on the boundary
belongs to the tensor product $V^{j_1}\otimes..\otimes V^{j_n}$.
 The only
constraint on physical states is that they should be globally
gauge invariant, i.e. $\SU(2)$ invariant, such that the possible
horizon states are the intertwiners (invariant tensors) between
the representations $V^{j_i}$ and the area is fixed at some value
$\aa$ \cite{etd2}.

Our model is different from the standard loop quantum  gravity
approach, which studies the classically induced boundary theory on
the black hole horizon (or generically any isolated horizon)
\cite{bhlqg}.
 Since the problem of identifing horizons within the quantum states of
geometry in full theory is yet unsolved, we model the black hole
as a region of the quantum space with a boundary (the horizon)
such that the only information about the geometry of the internal
region accessible to the external observer is information which
can be measured on the boundary. Our results can therefore be
applied to any closed surface.

In the simplest scenario, the black hole entropy calculation is
reduced to counting the number of distinct $\SU(2)$ invariant
states on the space of $2n$ qubits (spin-$\half$ states) with the
horizon area $\aa=a_{1/2}2n$. The ignorance of a particular
microstate makes the statistical state under consideration  be the
maximally mixed state $\rho$ on the space of intertwiners
$\hh_0=\Int (\C^2)^{\otimes 2n}$. Its orthogonal decomposition is
simply
\be
\rho=\frac{1}{N}\sum_i |\ii^i\9\6\ii^i|, \label{rho}
\ee
where $|\ii^i\9$  form n a basis of $\hh_0$and $N\equiv{\rm
dim}\,\hh_0$. A straightforward calculation results in the
Bekenstein-Hawking entropy and its logarithmic correction:
\be
S\equiv-\tr\rho\log\rho=\log N \sim 2n \log2-\ha3\log n.
\label{bh}
\ee

In a coarse-grained model of a black hole one considers horizon
states to be given by intertwiners between $m$ representations of
a fixed spin $s$, so the horizon area is $\aa=a_sm$. The
 space $\hh=(V^s)^{\otimes m}$ decomposes as
\be
\bigotimes^{m}\C^s\cong\bigoplus_{j=0}^{ms}\hh^j\equiv\bigoplus_{j}
V^j\otimes \S_{m,j}^s,\label{schur}
\ee
where $V^j$ is the irreducible spin-$j$ representation  of
$\SU(2)$, and $\S_{m,j}^s$ is the degeneracy subspace. From the
asymptotic form of the multiplicities $c_{mj}^s=\dim\S_{m,j}^s$ it
follows that
\be
S\sim m \log s-\ha3\log m.
\ee

\section{Entanglement and correlations}
Entanglement can be loosely defined as an exhibition of
stronger-than-classical correlations between the subsystems. 
Recently it
became one of the main resources of quantum information theory
\cite{per,nc}.
We are interested to find how much entanglement is contained in
arbitrary bipartite splittings of a horizon state. For a pure
state $|\Psi\9$ there is a unique measure of entanglement--- the
degree of entanglement which is the  entropy of either of the
reduced density matrices $\rho_\Psi^{A,B}$. I use here only one of
the measures of the mixed state entanglement, namely  the
entanglements of formation. It is possible to show that for the
states under consideration all measures of entanglement coincide
\cite{etd1}. The entanglement of formation is defined as follows.
A state $\rho$ can be decomposed as a convex combination of pure
states, $
\rho=\sum_\alpha w_\alpha |\Psi_\alpha\9\6\Psi_\alpha|$.
The entanglement of formation is the averaged degree of
entanglement of the pure states $|\Psi_\alpha\9$ (the von Neumann
entropy of their reduced density matrices) minimized over all
possible decompositions
\be
E_F(\rho)=\inf_{ \{ \Psi_\alpha\}}\sum_\alpha w_\alpha
S(\rho_\alpha).
\ee

 To simplify the notation consider again the qubit model of a black hole. Let the $2n$ qubits be divided
 into the groups of $2k\leq n$ and $2n-2k$ qubits. The corresponding Hilbert spaces  are $\hh_A\equiv(\C^2)^{\otimes {2k}}$ and
$\hh_B\equiv(\C^2)^{\otimes 2n-2k}$, respectively. Using the
decomposition of Eq.~(\ref{schur}) twice, the intertwiner space
can be decomposed as follows:
\be
\hh^0=V^0\otimes \S_{n,0}
=\bigoplus_{j=0}^k V^0_{(j)}\otimes (\S_{k,j}\otimes \S_{n-k,j}),
\label{schur2}
\ee
where $V^0_{(j)}$ is the singlet state in $V^{j}\otimes V^{j}$.
Hence the dimensionality of $\hh^0$ is related to the
multiplicities of the degeneracy subspaces  through
$N=c_{2n,0}=\sum_{j=0}^{k} c_{2k,j} c_{2n-2k,j}$.

The basis states of Alice and Bob are respectively labeled as
$|j,m, a_j\9$ and $|j,m, b_j\9$. Here $0\leq j\leq k (\leq n-k)$
and $-j\leq m\leq j$ have their usual meanings and the degeneracy
labels, $a_j$ and $b_j$, enumerate the different subspaces $V^j$

\be
|\ii^{a_j,b_j}\9\equiv\frac{1}{\sqrt{2j+1}}\sum_{m=-j}^j
(-1)^{j-m}|j,-m,a_j\9\otimes|j,m,b_j\9
\label{deco}
\ee
It is possible to show that the entanglement of formation of the
state $\rho$  is:
\be
E_F(\rho) =\frac{1}{N}\sum_{j=0}^{k}c_{2k,j}c_{2n-2k,j}\log(2j+1),
\label{ent1}
\ee
which is true for any spin $s\geq\half$. In the large $n$ limit
the case of Alice and Bob having $n$ qubits each is especially
interesting. In this case
\be
E_F(\rho|n:n)\sim\half\log n, \label{slog}
\ee
which is again true for any $s$ \cite{etd2}.

More generally, we find that for all bipartite splittings of the
spin network with sufficiently large number of qubits comprising
the smaller space, and for any $s\geq\half$, it is possible to
show that the the quantum mutual information between the black
hole horizon and its parts is  three times the entanglement
between the halves,
\be
I_\rho(A:B)\equiv S(\rho_A)+S(\rho_B)-S(\rho)\simeq
3S_E(\rho|A:B).
\ee
In particular,  if the ratio between the number of qubits is kept
fixed while $n$ is arbitrary, the logarithmic correction $\ha3\log
n$ asymptotically equals to $I_\rho(A:B)$, so the deviation of the
black hole entropy from its classical value equals to the total
amount of correlations between the halves of spin networks that
describe it \cite{etd2}. Hence  in a model where the black hole
horizon would be constructed out of independent uncorrelated
qubits, the entropy would scale  linearly in the number of qubits
$2n$. However, the requirement of invariance under $\SU(2)$
creates correlations between the horizon qubits, which are
revealed through the logarithmic correction $\ha3\log n$ to the
entropy law formula.

Returning to the qubit black hole it is interesting to note that a
fraction of unentangled states in Eq.~(\ref{deco}) when a pair of
qubits is segregated from the rest is $
s_0^{(2)}\sim\frac{1}{4}+\frac{3}{8n}$. It leads to an interesting
coincidence with the evaporation model \cite{bm} and allows to
speculate about corrections to it.

Moreover, using the relations between coarse-graining and partial
tracing it is possible to investigate the entanglement in
spin-networks \cite{etd3} and its possible role in the emergence
of classical geometry.

Finally, let me mention the ``information loss paradox"
\cite{t05}. While it is not obvious that the unitarity must
persist in the process of creation and evaporation of black holes,
consideration of the matter alone is not sufficient to
convincingly preserve it
\cite{rmp} . Entanglement between gravitational and matter
degrees of freedom offers the way to restore it. In the simplest
scenario initially the spacetime is approximately flat and the
matter is in some state $\rho$. We describe it by a state $\Phi$
that corresponds to a classical nearly Minkowski metric. The
evolution that ends in the black hole evaporation is unitary and
is schematically described as $
\Xi=U(\Phi\otimes\rho )U^{\dag}$,
where $\Xi$ is the final \emph{entangled} state of matter and
gravity. Reduced density operators  give predictions for the
gravitational background and the matter distribution on it. The
evolution of matter is obtained by tracing out the gravitational
degrees of freedom and is a completely positive non-unitary map
\cite{nc}. If we assume that the initial states are pure, then the
entropy of a reduced density operator is exactly the degree of
entanglement between matter and gravity, $E(\Xi)$. Hence, the
increase in the entropy of matter is not an expression of
information loss, but  a measure of the created entanglement, i.e.
\emph{redistribution} of information.

\acknowledgments

I thank Etera Livine for helping to introduce me to quantum
gravity and many enjoyable collaborations that provide the bulk of
this contribution.

\bigskip

\end{document}